\documentclass{article}
\usepackage{ijcai19}

\usepackage{times}
\usepackage{soul}
\usepackage{url}
\usepackage[hidelinks]{hyperref}
\usepackage[utf8]{inputenc}
\usepackage[small]{caption}
\usepackage{graphicx}
\usepackage{amsmath}
\usepackage{booktabs}
\usepackage{algorithm}
\usepackage{algorithmic}
  \usepackage{mathrsfs}
\usepackage{amssymb}
\usepackage{braket}
\usepackage[utf8]{inputenc}
\usepackage[english]{babel}
\newtheorem{theorem}{Theorem}

\DeclareMathOperator*{\argmin}{arg\,min}
\usepackage{dsfont}
\usepackage{multirow}
\usepackage{multirow}
\usepackage[table,xcdraw]{xcolor}
\usepackage{graphicx}
\usepackage{subcaption}
\usepackage{makecell}
\usepackage{graphicx}

\title{Wasserstein Collaborative Filtering for Item Cold-start Recommendation}

\author{
Yitong Meng$^1$
\and
Guangyong Chen$^2$\and
Benben Liao$^2$\and
Jun Guo$^3$\And
Weiwen Liu$^1$
\affiliations
$^1$the Chinese University of Hong Kong\\
$^2$Tencent Quantum Lab\\
$^3$Tsinghua-Berkeley Shenzhen Institute
\emails
\{ytmeng, wwliu\}@cse.cuhk.edu.hk,
\{gycchen,bliao\}@tencent.com,
eeguojun@outlook.com  
}

\begin{document}

\maketitle

\begin{abstract}
The item cold-start problem seriously limits the recommendation performance of Collaborative Filtering (CF) methods when new items have either none or very little interactions. To solve this issue, many modern Internet applications propose to predict a new item's interaction from the possessing contents. However, it is difficult to design and learn a map between the item's interaction history and the corresponding contents. In this paper, we apply the Wasserstein distance to address the item cold-start problem. Given item content information, we can calculate the similarity between the interacted items and cold-start ones, so that a user's preference on cold-start items can be inferred by minimizing the Wasserstein distance between the distributions over these two types of items. We further adopt the idea of CF and propose Wasserstein CF (WCF) to improve the recommendation performance on cold-start items. Experimental results demonstrate the superiority of WCF over state-of-the-art approaches.
\end{abstract}
\section{Introduction}
Recommender Systems (RSs) are extremely important nowadays to help users target their interested items among the massive amount of Internet information.
Collaborative Filtering (CF) has been the most widely used recommendation technique because of its promising performance demonstrated in the KDD Cup ~\cite{kddcup} and Netflix competition ~\cite{netflix}. For a specific user, CF recommends items according to the preference of users with similar rating history. However, CF is incapable of dealing with new items without any relevant interaction history, and thus seriously suffers from the cold-start problem.


Recent studies~\cite{saveski2014item,bar} show that the item cold-start problem can be effectively alleviated by considering the content information of items. The item content information is widely available in modern Internet applications, such as descriptions of products in Amazon, tags of movies in IMDb, images posted in Instagram, and so on.
Most successes in the item cold-start recommendation have employed a \textit{latent space sharing} model, assuming that an item shall preserve the same low dimensional representation for its interactions and content information.
These algorithms learn two individual projections,
through which an item's interaction history and content information are projected to the same point in the latent space. 
For a cold-start item having no interactions with users, the corresponding latent vector can be predicted from its content information. In such a way, the item-specific interactions among users can be computed based on the corresponding latent vectors.

Although the \textit{latent space sharing} methods achieve impressive success in recommending cold-start items in various recommendation scenarios, the designing of projection functions is difficult---which is the main focus of most current research---and sometimes must be specific to applications~\cite{van2013deep,wang2014improving}. Moreover, some types of content information require nontrivial projection methods, such as deep neural networks (DNNs), to learn the latent space of item contents. The high complexity of such projections further aggravates the training burden and possibly leads to over-fitting when data is small.


\begin{figure}[t]
	\centering
	\includegraphics[width=7.1cm]{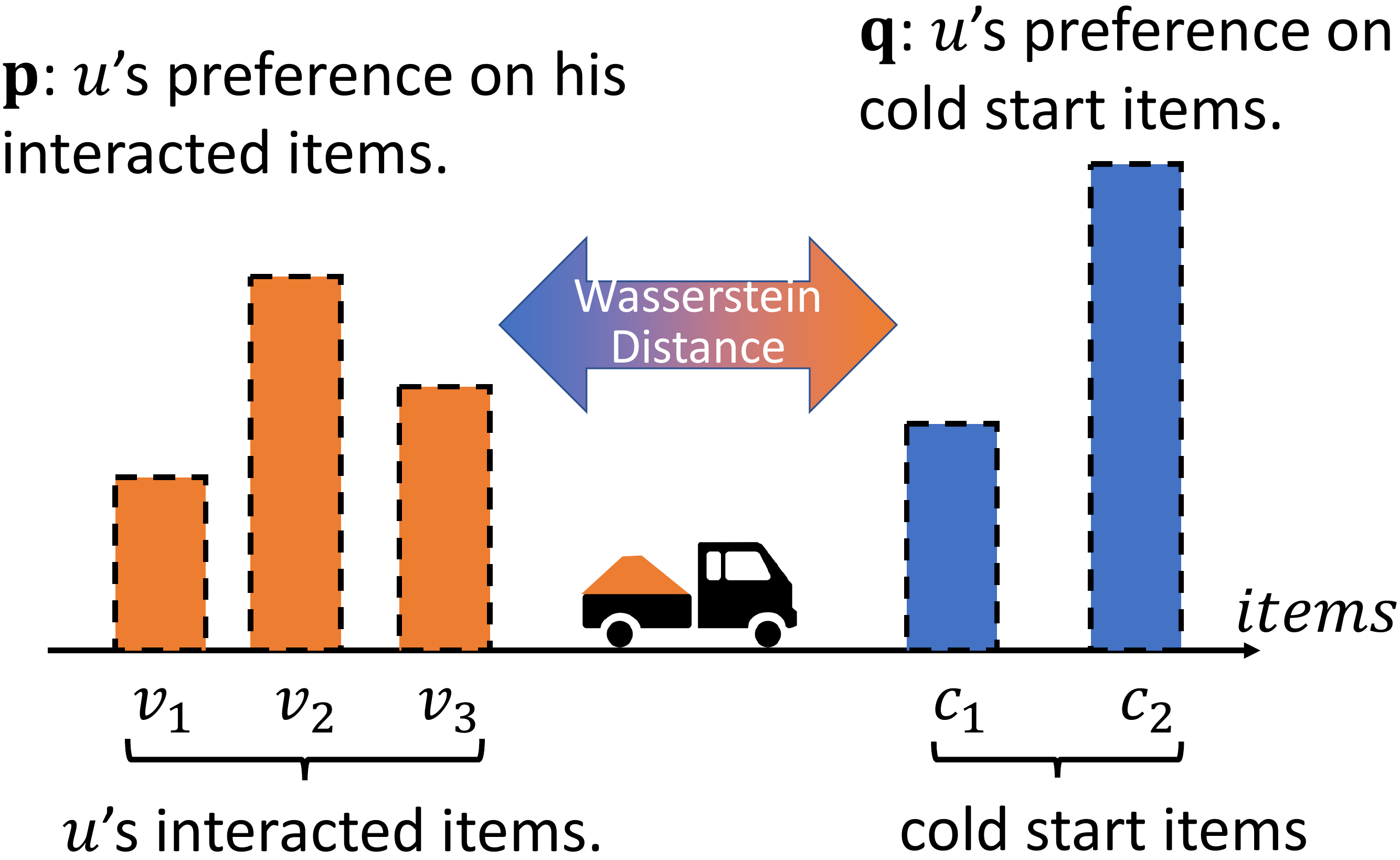}
	\caption{The Wasserstein distance between user $u$'s preference on the items that he interacted with and cold start items. We model a user's preference by a probabilistic distribution over items.}
	\label{fig:illu1}
	\vspace{-.5em}
\end{figure}
Instead of finding a common latent space of user-item interactions and item contents, we propose a Wasserstein Collaborative Filtering (WCF) approach to address the item cold-start problem. 
More specifically, we use the Wasserstein distance to measure the distance between a user's preference on his interacted items and cold start ones. 
As shown in Figure \ref{fig:illu1}, we define $\mathbf{p}$ as a user's preference on his interacted items and $\mathbf{q}$ his preference on cold-tart ones. 
The Wasserstein distance calculates the divergence between $\mathbf{p}$ and $\mathbf{q}$  given the similarity between two set of items, $\{v_1,v_2,v_3\}$ and $\{c_1,c_2\}$.  
Fortunately, item similarities can be extracted from the content information of items.
Then the user's preference on cold-start items can be easily solved by minimizing the Wasserstein distance. 
Inspired from the success achieved by CF, WCF collects preference information from many other users (collaborating) and assumes that the users' preference has a low-rank approximation optimized with respect to a Wasserstein distance.
As demonstrated empirically, WCF can further improve the performance of predicting users' preference on cold-start items.


To the best of our knowledge, no prior work has exploited optimal transport theory for collaborative filtering. Our contributions can be summarized as follows:
\begin{itemize}
\item To make full use of content information of items, we propose to measure the distances among users' preference by the Wasserstein distance.
\item we propose the Wasserstein filtering (WF) approach to infer a user's preference on cold-start items and further improve the performance by a CF technique.
\item Extensive experiments on three public real-world datasets demonstrate that our proposed WCF can significantly advance the state-of-the-art methods~\cite{wang2015collaborative,bar,saveski2014item,wang2011collaborative}.
\end{itemize}

\section{Backgrounds}
The most popular idea for performing CF is to represent both items and users by latent vectors so that ratings can be reconstructed from them. Although numerous instantiations~\cite{he2017neural,liang2018variational} of CF have been proposed in recent years, matrix factorization (MF)~\cite{mnih2007probabilistic,koren2009matrix} remains the most popular one due to its simplicity and effectiveness, and has been used for large scale recommendations of news~\cite{das2007google}, movies~\cite{koren2009matrix} and products~\cite{linden2003amazon}. 

Recent studies extend the MF framework for item cold-start recommendation by incorporating content information of items.
The majority of methods for item cold-start recommendation employ a latent space sharing model. For example, Saveski te al.~\shortcite{saveski2014item} and Barjasteh et al.~\shortcite{bar} propose to use MF as the prjection function for both interactions and item contents. LDA~\cite{wang2011collaborative}, CNN~\cite{kim2016convolutional}, DNN\cite{ebesu2017neural} , SDAE~\cite{wang2015collaborative,ying2016collaborative} and mDA~\cite{li2015deep} are proposed to learn the latent vectors of items from their textual contents. Van den Oord et al. ~\shortcite{van2013deep} and Wang et al. ~\shortcite{wang2014improving} propose to use CNN to learn the latent vectors of music from their audio signals. 

The Wasserstein distance, which originates from optimal transport theory~\cite{rubner1998metric,levina2001earth}, is a distance metric on probabilistic space and able to leverage the information on feature space. It has been successfully applied to many applications, such as computer vision ~\cite{arjovsky2017wasserstein} and natural language processing ~\cite{kusner2015word,huang2016supervised}, and is attracting more and more attention in academia. Recently, dictionary learning with a Wasserstein loss has been proposed for face recognition and topic modeling~\cite{rolet2016fast}. However, no prior works have applied optimal transport theory to collaborative filtering, especially the cold-start problem.
\section{The Proposed Model}
\subsection{Problem Definition}
\begin{figure}[t]
  \centering
    \includegraphics[width=8.1cm]{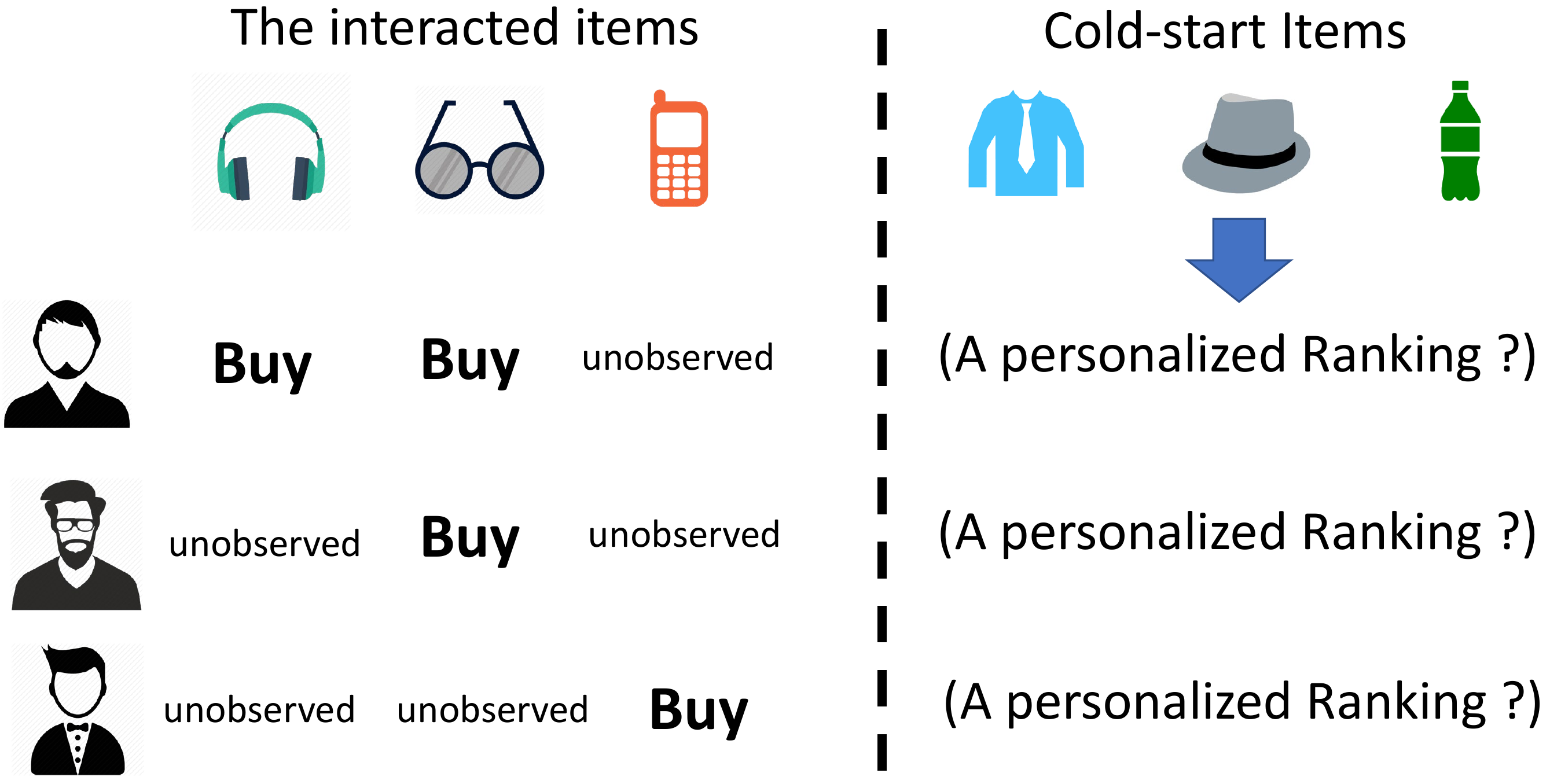}
    \caption{An illustration of problem definition.}
    \label{fig:prob}
      \vspace{-.5em}
\end{figure}

Let $ \mathcal{U} $ be a set of $m$ users, and $ \mathcal{V} $ a set of $n$ items which interact with the users in $ \mathcal{U}$. For simplicity, we call items in $ \mathcal{V} $ the \textit{interacted items}. 
The user-item interaction matrix is denoted by $ \mathbf{R} \in \mathbb{R}^{n\times m}$ with $\mathbf{R}_u$ as its $u$-th column vector, which denotes user $u$'s interactions on $\mathcal{V}$. $ \mathbf{R}_{vu} \geq 0$ is the interaction between user $u$ and item $v$, which can be a rating score, number of clicks or viewing times, etc. Usually $ \mathbf{R}$ is a partially observed matrix and highly sparse. Let $\mathcal{C} = \{c_1, c_2, ..., c_s\}$ be a set of cold-start items with no user interactions at all, and $\mathcal{C} \cap \mathcal{V} = \emptyset $. We are going to find a ranked list $ \mathcal{L}_u = (c_{u_1}, c_{u_2}, ..., c_{u_s})$ over $\mathcal{C}$ for a user $u \in \mathcal{U}$ such that $c_l$ ranks higher than $c_r$ in $ \mathcal{L}_u$ if $u$ prefers $c_l$ over $c_r$. Figure~\ref{fig:prob} briefly illustrates the problem definition.
\subsection{the Wasserstein Distance for User Preference}
\label{sec:WCB}
\noindent \textbf{User preference representation.}
Let $\Sigma_n=\{\mathbf{p}\in[0,1]^n \;|\; \braket{\mathbf{p},\mathds{1}}=1\}$ denote a $(n-1)$-dimensional simplex. 
We model a user’s preference over $\mathcal{V}$ as a probabilistic distribution $\mathbf{p}\in \Sigma_n$ on $ \mathcal{V}$, where $\mathbf{p}(v_l)>\mathbf{p}(v_r)$ if and only if $u$ prefers $v_l$ more than $v_r$. Similarly, we model $u$'s preference over $ \mathcal{C}$ as a distribution $\mathbf{q}\in \Sigma_s$ on $ \mathcal{C}$. 
Traditional methods for evaluating the distances among user preference in RSs is by the Euclidean distance or cosine value of their interaction vectors. However, such methods can not utilize content information and depend heavily on "commonly rated items". When two users interact with similar but disjoint items, such metrics tend to give a large distance value or are not applicable even if the real distance is small.

\noindent \textbf{User preference distance.}
The Wasserstein distance incorporates the item similarity to evaluate the distance of user preference $\mathbf{p}$ on $\mathcal{V}$ and $\mathbf{q}$ on $\mathcal{C}$.
Formally, 
the Wasserstein Distance first defines a polytope of transportation plans between $\mathbf{p}$ and $\mathbf{q}$ as follows:
\begin{align}
    U(\mathbf{p},\mathbf{q}) =\left \{ \mathbf{T}\in\mathbb{R}_{+}^{n\times s} \;|\;  \mathbf{T}\mathds{1}=\mathbf{p}, \mathbf{T}^T \mathds{1} =\mathbf{q} \right \}.
\end{align}
Then the Wasserstein distance between $\mathbf{p}$ and $\mathbf{q}$ is defined as:
\begin{align}\label{eq:1}
    W(\mathbf{p},\mathbf{q}) :=\min_{\mathbf{T}\in U(\mathbf{p},\mathbf{q})}  \braket{\mathbf{T},\mathbf{M}},
\end{align}
where $\mathbf{M}\in\mathbb{R}_{+}^{n\times s}$ is the \textit{cost matrix}, whose entry $\mathbf{M}_{ij}=d(v_i,c_j)$ evaluates the difference between items $v_i\in\mathcal{V} $ and $c_j\in \mathcal{C} $; $\mathbf{T}$ is called a \textit{transport plan}. For the specific definition of $\mathbf{M}$, please refer to Section ~\ref{data}. It is already shown that the Wasserstein distance is a metric~\cite{villani2008optimal}.

\noindent \textbf{Visualization.} To intuitively understand why the Wasserstein distance can measure the distances among users' preference, we introduce the concept of \textit{user utility}. A user's utility is maximized when his interactions ``matches" his preference. We also rename $\mathbf{M}$ as the \textit{utility cost matrix}, $\mathbf{T}$ as an \textit{exchange plan} for better understanding. 

\begin{figure}[t]
	\centering
	\includegraphics[width=7.5cm]{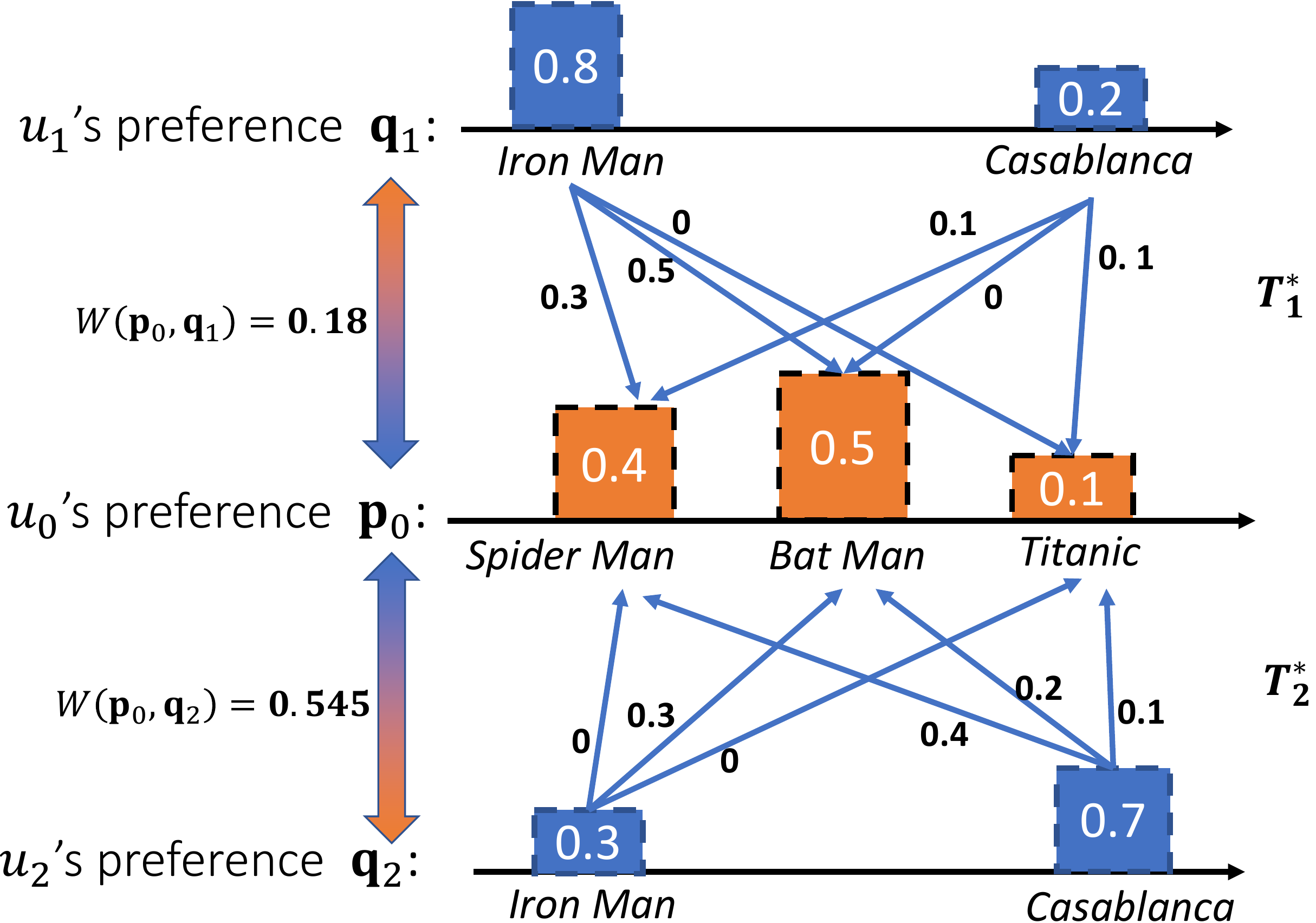}
	\caption{The components of the Wasserstein distance between $\mathbf{p}_0$ and $\mathbf{q}_1$, $\mathbf{q}_2$. The corresponding utility cost matrix $\mathbf{M}$ is in Table \ref{tab:M}. The one-way arrows $\mathbf{T}_1^*$ represent the optimal exchange plan between $\mathbf{p}_0$ and $\mathbf{q}_1$, labeled by the amount of exchange. $\mathbf{T}_2^*$ has a similar explanation. $\mathbf{T}_1^*$ and $\mathbf{T}_2^*$ exhibit that items tend to replace the ones similar to themselves. For example, $u_2$'s preference on \textit{Iron Man} (0.3) is entirely moved to \textit{Bat Man} because \textit{Iron Man} is most similar to \textit{Bat Man}. Therefore, users who enjoy similar items have a smaller distance between them, which matches the fact that $u_0$ and $u_1$ both prefers sci-fic movies while $u_2$ prefers romantic ones.}
	\label{fig:plan}
	\vspace{-.5em}
\end{figure}

For the ease of presentation, let's consider the example of $u_0$ and $u_1$ in Figure \ref{fig:plan}.
User $u_0$ spends one dollar on \{\textit{Spider Man}, \textit{Bat Man}, \textit{Titanic}\} according to his preference $\mathbf{p}_0$, i.e., \$0.4 for \textit{Spider Man}, \$0.5 for \textit{Bat Man} and \$0.1 for \textit{Titanic}. Similarly, user $u_1$ spends \$0.8 on \textit{Iron Man} and \$0.2 on \textit{Casablanca} according to his preference $\mathbf{q}_1$. Now, the utility of $u_0$ and $u_1$ is maximized because their interactions, i.e. how much money they spend, matches their preference. If $u_0$ and $u_1$ are asked to exchange all movies they consumed, how would the utility of the two users be affected? Due to the difference of movies, the exchange will incur a loss of user utility. The utility cost for exchanging similar movies is smaller than exchanging dissimilar ones. Table~\ref{tab:M} shows an example of a utility cost matrix $\mathbf{M}$, whose entry is the utility cost of exchanging two items. Based on $\mathbf{M}$, there exists an optimal exchange plan $\mathbf{T}$ that incurs the minimum utility loss for both users. Figure~\ref{fig:plan} shows the optimal exchange plan $\mathbf{T}_1^*$, which is the best choice for both $u_0$ and $u_1$. The corresponding utility loss is $\braket{\mathbf{T}_1^*,\mathbf{M}}=0.18$.
Thus, the smaller the minimum utility loss is, the closer the preference of the two users are.

\textbf{Remark 1:} The Wasserstein distance makes users who prefer similar items have a smaller distance, even if they have no commonly interacted items.
As shown in figure \ref{fig:plan}, we note that the optimal exchange plan tends to exchange similar items, because exchanging similar items loses less user utility. Consequently, the distance from $\mathbf{p}_0$ to $\mathbf{q}_1$ (0.18) is significantly smaller than to $\mathbf{q}_2$ (0.545). This agrees with the fact that $u_0$ and $u_1$ both like science fiction movies while $u_2$ likes romantic ones. However, Euclidiean distance and cosine can't measure the distance between $\mathbf{p}_0$ and $\mathbf{q}_1$ properly, because $u_0$ and $u_1$ have no commonly interacted items.
\begin{table}
\centering
\begin{tabular}{|c|c|c|c|}
\hline
  & \textit{Spider Man}  & \textit{Bat Man} & \textit{Titanic} \\
\hline
\textit{Iron man}      & 0.15  & 0.1 & 0.9 \\
\hline
\textit{Casablanca}       & 0.8  & 0.95 & 0.05  \\
\hline
\end{tabular}
\caption{The utility cost matrix $\mathbf{M}$ between \{\textit{Spider Man}, \textit{Bat Man}, \textit{Titanic}\} and \{\textit{Iron Man}, \textit{Casablanca}\}. \textit{Spider Man}, \textit{Bat Man} and \textit{Iron Man} are sci-fic, while \textit{Titanic} and \textit{Casablanca} are romantic. The utility costs among similar movies are smaller than those among dissimilar ones.}
\label{tab:M}
\vspace{-1em}
\end{table}
\noindent \textbf{Smoothed Wasserstein distance.} In many cases, the linear program defined in Equation (\ref{eq:1}) does not have a unique solution, and $W$ is not differentiable with respect to neither $\mathbf{p}$ nor $\mathbf{q}$. To address this problem, Cuturi ~\shortcite{cuturi2013sinkhorn} proposes an entropy regularizer to (\ref{eq:1}) and gets a new objective function as,
\begin{align}\label{eq:2}
    W_{\gamma}(\mathbf{p},\mathbf{q}) :=\min_{\mathbf{T}\in U(\mathbf{p},\mathbf{q})}  \braket{\mathbf{T},\mathbf{M}} - \gamma h(\mathbf{T}),
\end{align}
where $h$ is the entropy function $h(\mathbf{T}):=-\braket{\mathbf{T},\log\mathbf{T}}$.
Cuturi et al.\shortcite{cuturi2016smoothed} argues that when $\gamma>0$, equation (\ref{eq:2}) is differentiable with respect to either $\mathbf{p}$ or $\mathbf{q}$.
\subsection{Wasserstein Filtering}
A user $u$'s preference $\mathbf{q}_u$ on the cold-start items $\mathcal{C}$ can be inferred from his preference $\mathbf{p}_u$ on the interacted items $\mathcal{V}$, because we assume a user's taste is unchanged. This is a widely used assumption in recommendations~\cite{saveski2014item,bar}, which makes a user preserve the same latent vector for both the interacted items and cold-start ones.
In our model, this assumption is embodied as: $\mathbf{p}_u$ and $\mathbf{q}_u$ are the “same”.
Therefore, $\mathbf{q}_u$ can be estimated by minimizing its Wasserstein distance towards $\mathbf{p}_u$:
\begin{align}\label{WF}
\hat{\mathbf{q}}_u =\argmin_{\mathbf{q}_u \in \Sigma_s}W_\gamma(\mathbf{q}_u,\mathbf{p}_u).
\end{align}
$\mathcal{L}_u$ is therefore derived by ranking the values in $\hat{\mathbf{q}}_u$.
In practice, the ground truth of $u$'s preference $\mathbf{p}_u$ is also unobserved and is estimated by normalizing its interactions on $\mathcal{V}$, i.e. $\mathbf{p}_u \approx \frac{\mathbf{R}_u}{\braket{\mathbf{R}_u,\mathds{1}}}$, which is a commonly used method of user interest profiling in RSs.
\subsection{Wasserstein Collaborative Filtering}
However, the estimation $\mathbf{p}_u \approx \frac{\mathbf{R}_u}{\braket{\mathbf{R}_u,\mathds{1}}}$ contains noise because we treat unobserved interactions as ``zero'' preference and most interactions are unobserved.
As a result, inferring $\mathbf{q}$ for each user separately may not yield good performance. Therefore, we employ CF to address this issue by collecting preference information from many other users (collaborating) and assuming that the users’ preference has a low-rank approximation.
In particular, We perform CF by MF due to its simplicity and effectiveness: 
\begin{align}
\{{\mathbf{q}_u}\}_{u \in \mathcal{U}} \approx \mathbf{D}\mathbf{\Lambda}, \; s.t. \; \mathbf{D} \mathbf{\Lambda} \in (\Sigma_s)^m ,
\end{align}
where $\mathbf{D} \in \mathbb{R}^{s\times k}$ is the latent matrix of cold-start items, $\mathbf{\Lambda} \in \mathbb{R}^{k\times m}$ the latent matrix of users, and $k$ the latent dimension.
Therefore, the ultimate optimization objective of WCF is:
\begin{align}\label{cfobj}
\begin{split}
\mathbf{D},\mathbf{\Lambda}=&\argmin_{\mathbf{D} \in \mathbb{R}^{s\times k},\mathbf{\mathbf{\Lambda}} \in \mathbb{R}^{k\times m} } \sum_{u \in \mathcal{U}} W_\gamma(\mathbf{D}\mathbf{\Lambda}_u,\mathbf{p}_u),\\
& s.t. \; \mathbf{D}\mathbf{\Lambda} \in (\Sigma_s)^m .
\end{split}
\end{align}
\section{Optimization}
We use a block-coordinate descent approach  ~\cite{rolet2016fast} to optimize (\ref{cfobj}).
The optimization problems involved for Wasserstein Collaborative Filtering can be solved using dual problems whose objectives involve the Legendre-Fenchel conjugate of the smoothed Wasserstein distance~\cite{cuturi2016smoothed}. To abbreviate formulas, we define the following notation:
\begin{align}
H_\mathbf{p}:=\mathbf{q} \mapsto W_\gamma(\mathbf{p},\mathbf{q}),
\end{align}
where $p \in \Sigma_n$ and $q \in \Sigma_s$. The Legendre transform of the entropy regularized Wasserstein distance, as well as its gradient, can be computed in
closed form ~\cite{cuturi2016smoothed}:
\begin{align}
H_\mathbf{p}^*(g):=\gamma(h(\mathbf{p}),\braket{\mathbf{p}, \log \mathbf{K} \alpha}),
\end{align}
\begin{align}
\nabla H_\mathbf{p}^*(g):= \alpha \odot \mathbf{K}^T \frac{\mathbf{p}}{\mathbf{K}\alpha}.
\end{align}
where $g \in \mathbb{R}^s$ is the conjugate variable, $\mathbf{K}:=e^{-\mathbf{M}/\gamma}$ and $\alpha:=e^{g/\gamma}$.

$\mathbf{D}$ and $\mathbf{\Lambda}$ in (\ref{cfobj}) can be solved in a block-coordinate descent manner:
\subsubsection{$\mathbf{\Lambda}$ step}
Consider $\mathbf{D}$ is fixed, and our goal is to compute:
\begin{align}\label{eq:5}
\begin{split}
\mathbf{\Lambda}^*=&\argmin_{\mathbf{\Lambda} \in \mathbb{R}^{k \times m}} \sum_{u\in \mathcal{U}} W_\gamma(\mathbf{D}\mathbf{\Lambda}_u,\mathbf{p}_u),\\
& s.t. \; \mathbf{D}\mathbf{\Lambda} \in (\Sigma_s)^m.
\end{split}
\end{align}
\begin{theorem}
Let $\mathbf{\Lambda}^*$ be a solution of Problem (\ref{eq:5}). $\mathbf{\Lambda}^*$ satisfies $\mathbf{D}\mathbf{\Lambda}_u^*=\nabla H_{\mathbf{p}_u}^*(g_u^*)$ for $u \in \mathcal{U}$, with 
\begin{align}\label{eq:6}
g_u^* \in \argmin_{g \in \mathbb{R}^s} H_{\mathbf{p}_u}^*(g), \,s.t. \,\mathbf{D}^Tg=0.
\end{align}
Moreover, if $\mathbf{D}$ is full-rank this solution is unique
~\cite{rolet2016fast}.\end{theorem}
Equation (\ref{eq:6}) can be solved with a projected gradient descent approach and then $\mathbf{\Lambda}^*$ is recovered by solving the linear equation $\mathbf{D}\mathbf{\Lambda}^*=\{\nabla H_{\mathbf{p}_u}^*(g_u^*)\}_{u \in \mathcal{U}}$~\cite{rolet2016fast}.

\subsubsection{$\mathbf{D}$ step}
Consider $\mathbf{\Lambda}$ is fixed, and our goal is to compute:
\begin{align}\label{eq:7}
\begin{split}
\mathbf{D}^*=&\argmin_{\mathbf{D} \in \mathbb{R}^{s \times k}} \sum_{u \in \mathcal{U}} W_\gamma(\mathbf{D}\mathbf{\Lambda}_u,\mathbf{p}_u),\\
& s.t. \; \mathbf{D}\mathbf{\Lambda} \in (\Sigma_s)^m.
\end{split}
\end{align}
\begin{theorem}
Let $\mathbf{D}^*$ be a solution of problem (\ref{eq:7}). $\mathbf{D}^*$ satisfies $\mathbf{D}^*\mathbf{\Lambda}=\{\nabla H_{\mathbf{p}_u}^*(g_u^*)\}_{u \in \mathcal{U}}$, with 
\begin{align}\label{eq:8}
G^* \in \argmin_{G \in \mathbb{R}^{s\times m}} \sum_{u \in \mathcal{U}} H_{\mathbf{p}_u}^*(G_u),\,s.t.\, G \mathbf{D}^T=0.
\end{align}
Moreover, if $\mathbf{\Lambda}$ is full-rank this solution is unique
~\cite{rolet2016fast}.\end{theorem}
Equation (\ref{eq:8}) can be solved with a projected gradient descent approach and then $\mathbf{D}^*$ is recovered by solving the linear equation $\mathbf{D}^*\mathbf{\Lambda}=\{\nabla H_{\mathbf{p}_u}^*(G_u^*)\}_{u \in \mathcal{U}}$~\cite{rolet2016fast}.

\section{Experiments}
In this section, we conduct extensive experiments to demonstrate the merits and advantages of the proposed algorithm. In particular, we focus on exploring several fundamental questions:
\begin{itemize}
\item How does the proposed model affect the performance of recommending cold-start items to existing users, compared to the state-of-the-art algorithms?
\item  How does the ratio of train to test affect the performance of the proposed algorithm?
\item How does CF affect the performance of the proposed algorithm?
\item  How does the latent dimension $k$ affect the performance of the proposed algorithm?
\end{itemize}
\subsection{Datasets}
\label{data}
We conduct our experiments on three well-known datasets: MovieLens-100k, MovieLens-1M and MovieLens-10M\footnote{The three datasets are available at \url{https://grouplens.org/datasets/movielens/}}. These are user-movie ratings collected from a movie recommendation service website: movielens.org. 
\begin{table}
\small
\vspace{-1em}
\centering
\begin{tabular}{llll}
\hline
datasets  & \makecell{Movie\\Lens-100k} & \makecell{Movie\\Lens-1M} & \makecell{Movie\\Lens-10M} \\
\hline
\# of users       & 942  & 6038   & 71554 \\
\# of items       & 1228  & 3395 &    9728  \\
\# of interactions    & 48216  & 573360  & 5603219  \\
\% of interactions  & 4.16\%  & 2.79\%  & 0.80\%\\
\hline
\end{tabular}
\caption{statistics of MovieLens-100k, MovieLens-1M and MovieLens-10M after preprocessing}
\label{tab:stat}
\vspace{-1em}
\end{table}
Tag-genomes~\cite{vig2012tag} are used as the content information of movies. A tag-genome of a movie is a vector.  Each value in this vector is a relevance score on a continuous scale from 0 to 1, representing the relevance of a tag to a movie. In the MovieLens tag-genome dataset\footnote{\url{https://grouplens.org/datasets/movielens/tag-genome/}}, tag relevance values are provided for 9,734 movies and 1,128 tags.

Following the data processing in~\cite{liang2018variational,wang2015collaborative,ebesu2017neural}, we binarize the explicit data by keeping ratings of four or higher and interpret them as implicit feedback. We also remove movies without tag-genomes and users with no interactions.
Similar to ~\cite{yao2018judging}, we compute the cosine value of the tag-genomes of two movies as the similarity between them. Practically, we define the cost matrix by $\mathbf{M}_{ij}:=1-Sim(v_i,v_j)$, where $Sim(v_i,v_j)\in [0,1]$ is the similarity of $v_i$ and $v_j$. Several other mappings from similarity to utility costs are explored but do not show significant improvement over this simple approach. Thus, they are not reported.
Table \ref{tab:stat} shows the statistics of the datasets after processing.

\subsection{Evaluation Metric}
We use three widely used ranking-based metrics: mean average precision (MAP), the truncated normalized discounted cumulative gain (NDCG@R) and Recall@R. For each user,
all metrics compare the predicted rank of the held out items with their true rank. 
Formally, we define $\mathcal{L}_u(r)$ as the item at rank $r$ in the ranked list $\mathcal{L}_u$ over $\mathcal{C}$, $I_u \in \mathcal{C}$ as the set of held-out items that user $u$ clicks on, $\mathds{1}[\cdot]$ the indicator function. Then we can compute the Average Precision (AP) of user $u$ as:
\begin{align}
AP(u,\mathcal{L}_u):=\frac{\sum_r \mathds{1}[\mathcal{L}_u(r) \in I_u] \times Precision@r}{\sum_r \mathds{1}[\mathcal{L}_u(r)\in I_u]}.
\end{align}
Where $Precision@r=\frac{\sum_{i=1}^r \mathds{1}[\mathcal{L}_u(i)\in I_u] }{r}$. MAP is the average of AP over all the users.
NDCG@R is defined as:
\begin{align}
NDCG@R(u,\mathcal{L}_u):=Z_R \sum_{r=1}^R \frac{2^{\mathds{1}[\mathcal{L}_u(r)\in I_u -1 ]}}{\log r +1}.
\end{align}
where R is called the scope, which means the number
of top-ranked items presented to the user and $Z_R$ is chosen
such that the perfect ranking has an NDCG@R value of 1.
Recall@R for user $u$ is: 
\begin{align}
Recall@R(u,\mathcal{L}_u):=\frac{\sum_{r=1}^{R}\mathds{1}[\mathcal{L}_u(r)\in I_u]}{|I_u|}.
\end{align}
In our experiment, scope R is selected to be 20, which is a normal choice in most recommender system literature. The average of NDCG@20 over all users is presented as the final result, and so is Recall@20.
\subsection{Experimental Setup}
We split each dataset into two subsets, a training set that used for training the model and a test set that used for evaluating it. 
More specifically, in order to create cold-start scenario, the train/test sets are split in the following manner: Randomly split the items into two subsets, an `interacted items' set $\mathcal{V}$ and a `cold start items' set $\mathcal{C}$; Then the training set is all the interactions on $\mathcal{V}$ and the testing set all the interactions on $\mathcal{C}$.

The ratio of training data to testing data is selected to be 3:1, 1:1 and 1:3. We apply cross-validation for each ratio. For example, when train:test=3:1, we partition the data into 4 equal subsets and then use one for testing and the rest for training. We repeat the experiment for each of the partitions.
\subsection{Baselines}
To evaluate the performance of our proposed WCF algorithm, we consider several state-of-the-art baseline approaches. 
In particular, we choose the following baseline algorithms:
\begin{itemize}
\item Content Based Filtering (CBF) ~\cite{saveski2014item}: This algorithm builds a profile for each user based on the properties of the past user’s preferred items. The items whose profiles are closest to the user's profile are recommended to the user. 
\item Local Collective Embedding (LCE) ~\cite{saveski2014item}: This algorithm factorizes the interaction matrix and content matrix simultaneously, making them sharing a common item latent matrix. 
\item  Decoupled Recommendation (DecRec)~\cite{bar}: This algorithm firstly factorizes user-item matrix and then factorizes item similarity matrix.
\item Collaborative Topic Regression (CTR)~\cite{wang2011collaborative}: This model
performs topic modeling on the item content matrix and collaborative filtering on the interaction matrix simultaneously.
\item Collaborative Deep Learning (CDL)~\cite{wang2015collaborative}: This algorithm applies MF on the interaction matrix and SDAE on the content information of items. 
\end{itemize}
Table \ref{tab:para} reports the best performing values of parameters of all implemented models found by grid search. 
\begin{table}[t]
\vspace{-0.5em}
\small
\centering
\begin{tabular}{c|c}
\hline
Models&Parameter settings  \\
\hline
 CBF& --- \\
 \hline
 LCE& $\alpha=0.5, \lambda=0.5, \beta=0.25, k=500.$\\
 \hline
 DecRec& $k=100. $\\
 \hline
 CTR& $\lambda_u=0.1,\lambda_v=10, a=1, b=0.01, k=50.$\\
 \hline
  CDL&  \makecell{$\lambda_u=0.1, \lambda_v=10, \lambda_w=0.001, a=1,$ \\$b=0.01, k=50.$}\\
  \hline
 WF& $\gamma=0.05.$\\
 \hline
WCF& $\gamma=0.05, k=30.$\\
\hline
\end{tabular}
\caption{Parameter settings of baselines.}
\label{tab:para}
\vspace{-1em}
\end{table}
 \begin{figure}[b]
	\centering
	\includegraphics[width=6.3cm]{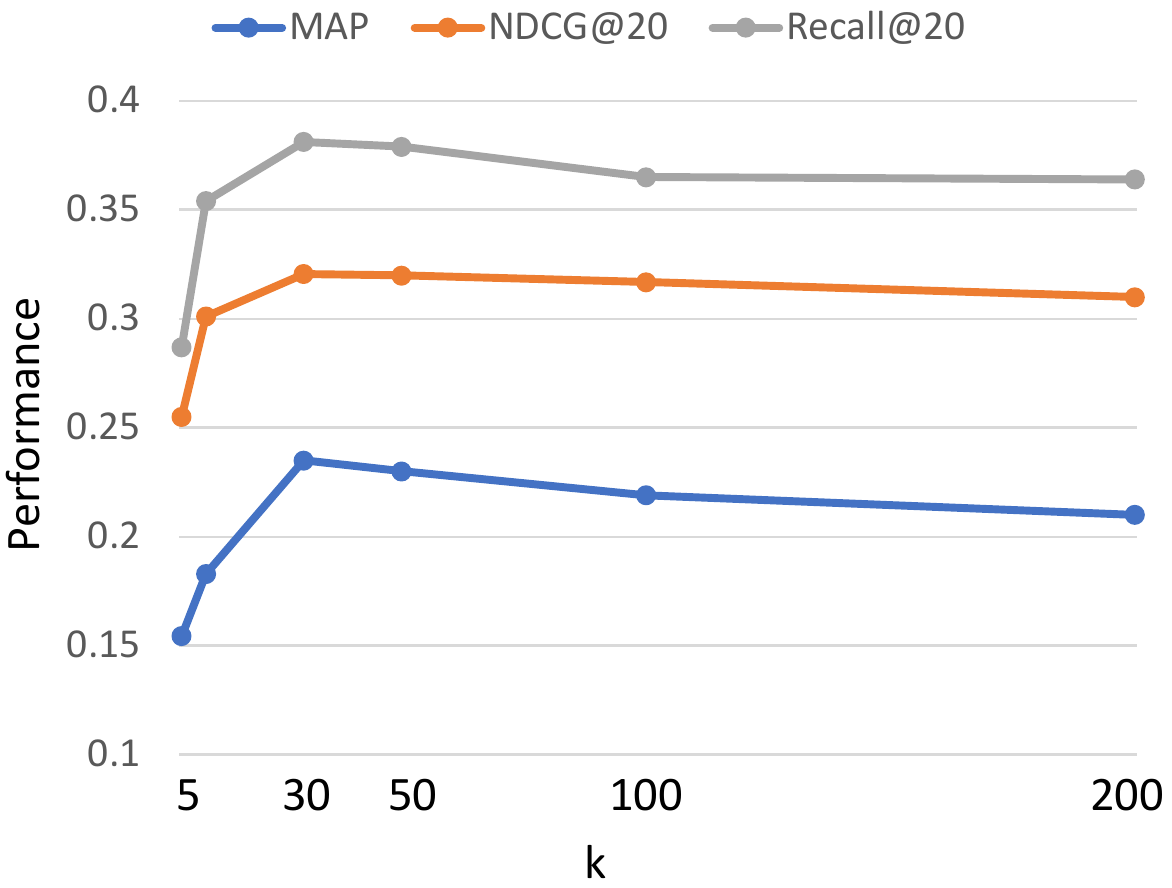}
	\caption{The performance of WCF with different values of latent dimension $k$ on the Movielens-1M data set ($\gamma=0.05$, train:test=3:1).}
	\label{fig:para}
\end{figure}
\begin{figure*}[t]
	\centering
	\includegraphics[width=16.9cm]{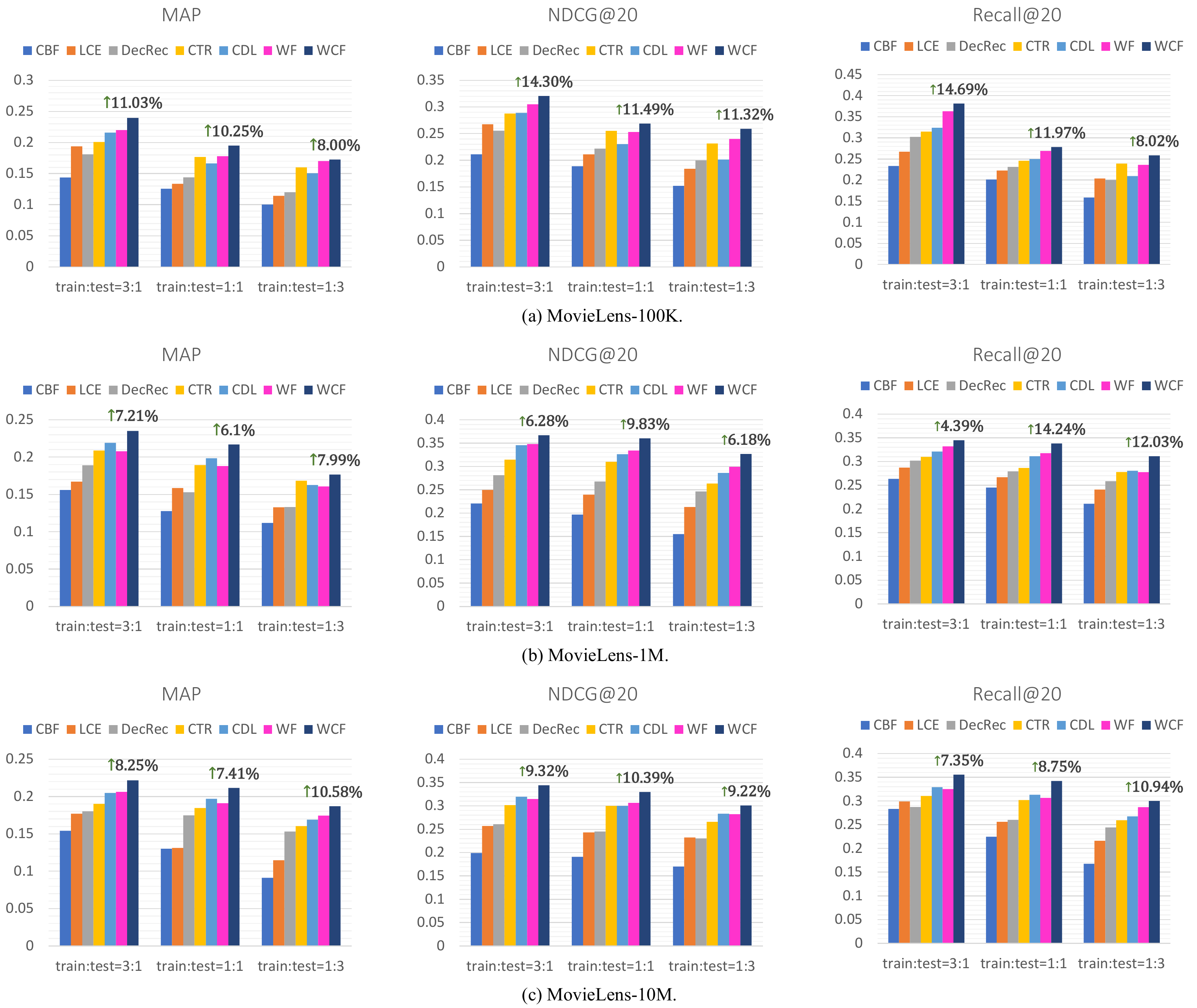}
	\caption{Performance of various baselines on MovieLens-100K, -1M and -10M. The numbers beside the green arrows represent the improvement over the best baselines.}
	\label{fig:result}
	\vspace{-1.5em}
\end{figure*}
\subsection{Results}

By the results shown in Figure \ref{fig:para} and \ref{fig:result}, we are able to answer the questions presented at the beginning of this section.

 \textit{\bf Prediction accuracy of cold start items recommendation}: 
 WCF performs best among all baselines and is significantly better than the second place in all settings.
 Moreover, WF already performs better or even than other baselines in most cases.
 This demonstrates the Wasserstein Distance is a powerful method to infer a user's preference on cold-start items.
 
 \textit{\bf Robustness to limited training data}: Our algorithm performs best across all train/test ratios. WCF still achieves a good performance even when the train set is only one-quarter of the original data, while the performance of other baselines decreases relatively faster when training data become small. 

 \textit{\bf The effect of CF in WCF}: In order to evaluate the benefit of the CF technique in WCF, we compare WCF to WF, which is a truncated version of WCF without matrix factorization. As shown in Figure \ref{fig:result}, the matrix factorization technique gives a boost of nearly 10\% to WCF over WF. This again demonstrates that the users' preference is in a low-dimension space and collaborative filtering is a powerful method to utilize this structure.

 \textbf{Hyperparameter analysis}: It can be found that the latent dimension $k$ controls the complexity of WCF. The behavior of $k$ is plotted in figure \ref{fig:para}. When k is small, the model is too simple to capture the patterns in the data, resulting in under-fitting. When $k$ increases, the performance goes up and peaks at $k=30$. Afterward, the performance slightly decreases due to over-fitting. However, such a decrease is extremely slow, which is possibly due to the robustness of the Wasserstein distance.

\section{Conclusion}
We use the Wasserstein distance to infer a user's preference on the cold-start items based on his historical interactions and  item contents. Collaborative filtering is applied during this process to further improve the prediction accuracy.
Empirical results show that WCF can significantly advance the state-of-the-art methods and robust to small training data.




\appendix


\clearpage
\bibliographystyle{named}
\small
\bibliography{ijcai19}

\end{document}